\documentclass[a4paper,11pt]{article}

\pdfoutput=1

\usepackage{jheppub}
\usepackage{amsfonts}
\usepackage{amsmath}
\usepackage{amssymb}
\usepackage{graphicx,color}
\usepackage{float}
\usepackage{hyperref}
\usepackage{subfigure}

\usepackage{xcolor}
\setcitestyle{square, numbers, comma, sort&compress}

\title{\boldmath Quantum phase transitions in a bidimensional
  $O(N) \times {\mathbb{Z}_2}$ scalar field model }

\author[a]{Gustavo O. Heymans}    
\author[b]{, Marcus Benghi Pinto}    
\author[c]{and Rudnei O. Ramos}

\affiliation[a]{Centro Brasileiro de Pesquisas F\'{\i}sicas, 22290-180 Rio de Janeiro, RJ, Brazil}
\affiliation[b]{Departamento de F\'{\i}sica, Universidade Federal de Santa
  Catarina, 88040-900 Florian\'{o}polis, SC, Brazil}
\affiliation[c]{Departamento de F\'{\i}sica Te\'orica,
  Universidade do Estado do Rio de Janeiro,
  20550-013 Rio de Janeiro, RJ, Brazil}

\emailAdd{olegario@cbpf.br}
\emailAdd{marcus.benghi@ufsc.br}
\emailAdd{rudnei@uerj.br}

\abstract{ We analyze the possible quantum phase transition patterns
  occurring within the $O(N) \times {\mathbb{Z}_2}$ scalar multi-field
  model at vanishing temperatures in $(1+1)$-dimensions. The physical
  masses associated with the two coupled scalar sectors are evaluated
  using the loop approximation up to second order. We observe that in
  the strong coupling regime,  the breaking $O(N) \times
  {\mathbb{Z}_2} \to O(N)$, which is allowed by the
  Mermin-Wagner-Hohenberg-Coleman theorem, can take place through a
  second-order phase transition. In order to satisfy this no-go
  theorem, the $O(N)$ sector must have a finite mass gap for all
  coupling values, such that conformality is never attained, in
  opposition to what happens in the simpler ${\mathbb{Z}_2}$
  version. Our evaluations also show that the sign of the
  interaction between the two different fields alters the transition
  pattern in a significant way.  These results may be relevant to
  describe the quantum phase transitions taking place in cold linear
  systems with competing order parameters. At the same time the
  super-renormalizable model proposed here can turn out to be useful
  as a prototype to test resummation techniques as well as
  non-perturbative methods.  } 
 

\arxivnumber{ 2205.04912 [hep-th]}

\begin{document} 
\maketitle


\section{Introduction}

The  $\lambda\phi^4$ scalar field model with ${\mathbb{Z}_2}$ symmetry
in $(1+1)$-dimensions ($\phi^4_2$) is a rather simple  quantum field
theory model, yet complete enough to illustrate and help to understand
many important aspects of much more complicated quantum field theory
models.  {}For this reason, it has been extensively considered in the
literature (for a far from complete list of references see, e.g.,
refs.~\cite{Hauser:1994mb,Rychkov:2014eea,Serone:2018gjo,Serone:2019szm,Romatschke:2019wxc,Romatschke:2019rjk,Kadoh:2018tis,Bronzin:2018tqz,
  Heymans:2021rqo}). The $\phi^4_2$ model describes a simple
non-integrable super-renormalizable theory which can display rich
phase transition patterns. When the original square mass parameter of
the model, $m^2$, is positive, the model has a mass gap and remains
invariant under the $\mathbb{Z}_2$ symmetry,  as far as one remains
within the weak coupling regime. As the strength of the interaction
increases, the mass gap decreases until the symmetry gets ultimately
broken through a second-order quantum phase transition when a critical
coupling value is reached~\cite{Simon:1973yz,Chang:1976ek}. Since the
model's $\beta$-function vanishes at all orders in perturbation
theory, it  represents a conformal field theory  at the critical
coupling where it becomes gapless. It then lies within the same
universality class as the bidimensional Ising model.  These physically
appealing characteristics,  combined with its simplicity, turn the
model into a perfect  framework to test how accurately different
nonperturbative techniques describe critical parameters associated
with the phase transitions. 

{}From the symmetry breaking point of view, it should be recalled that
Coleman's theorem~\cite{Coleman:1973ci} prevents the breaking of a
continuous symmetry in $(1+1)$-dimensions at the quantum level. When
extended to finite temperatures we then refer to the
Mermin-Wagner-Hohenberg theorem~\cite{Mermin:1966fe,Hohenberg:1967zz},
which states that at finite temperatures, no continuous spontaneous
symmetry breaking can occur for $d\leq 3$,  where $d$ is the spacetime
dimension. {}Finally, still in the context of  low-dimensional
systems, Landau's theorem~\cite{Landau:1980mil} prevents symmetry
breaking (be it a continuous or a discrete one in nature) in one-space
dimension and at finite temperatures. It is an important consistency
check to have these no-go theorems observed in these low-dimensional
systems. At the same time, scalar systems with a large symmetry, such
as $O(N_\phi) \times O(N_\chi)$ play an important role in a variety of
physical situations, including anisotropic antiferromagnets in an
external magnetic
field~\cite{Kosterlitz:1976zza,Calabrese:2002bm,Eichhorn}, high-$T_c$
superconductors~\cite{Demler:2004zz} and possibly even have a role in
understanding aspects of extensions of the Standard Model of
elementary
particles~\cite{Meade:2018saz,Baldes:2018nel,Matsedonskyi:2020mlz,Matsedonskyi:2020kuy,Bajc:2020yvd,Chaudhuri:2020xxb,Chaudhuri:2021dsq,Niemi:2021qvp,Ramazanov:2021eya}. The
distinctive transition patterns displayed by these type of models are
due to the competition between  two coupled fields described by a
classical interaction potential of the form~\cite{Kosterlitz:1976zza} 
\begin{equation}
V_{\rm int}(\phi,\chi) = \frac{\lambda_\phi}{4!} \phi^4 +
\frac{\lambda_\chi}{4!} \chi^4 + \frac{\lambda}{4} \phi^2 \chi^2,
\label{Vintphichi}
\end{equation}
where $\phi$ and $\chi$ can be $N_{\phi (\chi)}$-component fields,
$\phi \equiv (\phi_1, \ldots, \phi_{N_\phi})$ and $\chi \equiv
(\chi_1, \ldots, \chi_{N_\chi})$.  To avoid a runaway in the
interacting potential, eq.~(\ref{Vintphichi}), one needs to enforce
boundness from below, which requires the couplings to satisfy
$\lambda_\phi,\,\lambda_\chi \ge 0$ and, if $\lambda < 0$, one  needs
to further impose  $\lambda_\phi \lambda_\chi > 9 \lambda^2$. The fact
that the interspecies coupling, $\lambda$, may be negative can lead to
some exotic transition patterns such as {\it inverse symmetry
  breaking} as well as {\it symmetry non-restoration}, which were
observed in a seminal paper by  Weinberg~\cite{Weinberg:1974hy} (in
$(3+1)$-dimensions and at finite temperatures). 

As far as planar systems are concerned and in view of the
Mermin-Wagner-Hohenberg-Coleman (MWHC) theorem, an interesting
possibility has been recently explored in ref.~\cite{Chai:2021djc}
(see also refs.~\cite{Chai:2020zgq,Chai:2020onq} for related works),
where the authors analyze the possible phase transition pattern
$O(N_\phi) \times {\mathbb{Z}_2} \to O(N)$ occurring in hot planar
coupled systems. In the present paper, our aim is to explore the
possible (quantum) phase transition patterns exhibited by the same
model at vanishing temperatures and in $(1+1)$-dimensions.  The model studied in this paper can also be seen as reminiscent of a two-dimensional $XY$-Ising model of statistical mechanics (see, e.g., ref.~\cite{delfino} and references therein). The $O(N_\phi) \times {\mathbb{Z}_2}$ model can also be seen as an extension to lower
dimensions, and to the quantum domain, of  many previous studies on
similar coupled scalar field
models~\cite{Mohapatra:1979qt,Klimenko:1988mb,Bimonte:1995xs,AmelinoCamelia:1996hw,Orloff:1996yn,Roos:1995vm,Jansen:1998rj,Bimonte:1999tw,Pinto:1999pg,Pinto:2005ey,Pinto:2006cb,Farias:2021ult}
which were carried out in the presence of  a heat bath  in
$(3+1)$-dimensions. 

The paper is organized as follows. In the next section we review the
simplest one-component scalar field case with ${\mathbb{Z}_2}$
symmetry. Then, in Sec.~\ref{section3}, we present the $O(N_\phi)
\times {\mathbb{Z}_2}$ model, corresponding to two coupled scalar
fields for the general $N_\phi \ge 2$ case.  In the same section we
evaluate the physical squared masses up to two-loop   which represents
the first non trivial contribution.  Numerical results associated with
the phase transition patterns are generated and discussed in
Sec.~\ref{section4}. Our conclusions are then presented in
Sec.~\ref{conclusions}. {}For completeness, some technical details
related to the original evaluation of two-loop diagrams with different
masses are presented in an Appendix.

\section{Warm up: the ${\mathbb{Z}_2}$ model}
\label{section2}

To make the paper self-contained, let us start by reviewing some well
known results regarding the ${\mathbb{Z}_2}$ model, which is described
by the Lagrangian density (in Euclidean space)
\begin{eqnarray}
   \mathcal{L} =\frac{1}{2}(\partial_{\mu}\chi)^2 +
   \frac{1}{2}m^2\chi^2 + \frac{\lambda}{4!}\chi^4 \, ,
\label{Z2subcritical}   
\end{eqnarray}
where, in order to investigate the symmetric phase, we take $m^2 >0$.
In (1+1)-dimensions the theory is super-renormalizable and $\lambda$
has canonical dimensions $[\lambda] = 2$. In this particular case the
coupling is finite and, hence, the renormalization group
$\beta$-function is just $\beta_\lambda = 0$ at all perturbative
orders. Regarding the two-point function, the only primitive
divergence stems from tadpole (direct) contributions,  which do not
depend on the external momentum, implying that no wave-function
renormalization is needed. Thus, the mass anomalous dimension reads
\cite{Serone:2019szm}
\begin{equation}
\beta_{m^2} = - \frac{\lambda}{4 \pi} \,.
\label{betam}
\end{equation}

\subsection{The physical mass}

Let us now evaluate the physical square mass for the model
(\ref{Z2subcritical}) by  considering two regions: the (subcritical)
region, where the model is  ${\mathbb{Z}_2}$ symmetric, and the
(supercritical) region, where this symmetry is broken.

The subcritical region ($\lambda < \lambda_c$) is described directly
by the Lagrangian density represented by eq.~(\ref {Z2subcritical}),
from which one may evaluate the physical square mass. {}Following most
applications (e.g.,
refs.~\cite{Serone:2018gjo,Romatschke:2019rjk,Heymans:2021rqo}), we
will evaluate the pole mass which means that all self-energies are to
be evaluated on mass-shell ($p^2= - m^2$ in Euclidean space). In this
case, a semiclassical  expansion in loops, to two-loop order yields
\begin{eqnarray}
M^2 &=& m^2 + \frac{\lambda}{8\pi}L_m - \frac{\lambda^2}{384  m^2} -
\frac{\lambda^2}{64 \pi^2 m^2}L_m ,
\label{massaZ2sub1}
\end{eqnarray}
where $L_m$ is 
\begin{equation}
    L_m = \ln \frac{\mu^2}{m^2},
\end{equation}
with $\mu$ representing the $\overline{\rm MS}$ dimensional
regularization scale.

After the dynamical symmetry breaking, which is defined by a critical
coupling $\lambda_c$ at which $M^2(\lambda=\lambda_c)=0$, a
nonvanishing vacuum expectation value (VEV)  for the field
$\overline{\chi}$, develops.  One can then perform the usual shift
around the new developed VEV: $\chi  \to \chi^\prime = \chi -
\overline{\chi}$. After that, the supercritical region ($\lambda >
\lambda_c)$ is   described by the following (Euclidean) Lagrangian
density
\begin{equation}
       \mathcal{L}' =\frac{1}{2}(\partial_{\mu}\chi^\prime)^2  +
       \frac{1}{2}\Omega^2(\chi^\prime)^2 +
       \frac{\lambda}{4!}(\chi^\prime)^4 +
       \frac{\lambda}{3!}\overline{\chi}(\chi^\prime)^3\,,
\label{Lprime}
\end{equation}
where
\begin{equation}
    \Omega^2 = m^2 + \frac{\lambda}{2}\overline{\chi}^2\,.
\end{equation}
Note that linear terms do not appear in eq.~(\ref {Lprime}) since they
all cancel as expected. Hence, at the classical (tree-)level,
$\overline{\chi}$ is fixed by
\begin{equation}
m^2\overline{\chi} + \frac{\lambda\overline{\chi}^3}{6}= 0,
\end{equation}
which  has $\overline{\chi}=0$ as its only real solution when $m^2>0$
. As expected quantum corrections will produce a non-vanishing
$\overline{\chi}$ beyond a critical coupling value as we show next.

The contributions up to two-loop order for both the physical square
mass and  $\Gamma^{(1)}$ now include terms that can be constructed
from the tri-linear vertex in eq.~(\ref{Lprime}). In this case, the
diagrams contributing to the physical square mass  and to the tadpole
equation $\Gamma^{(1)}$ are, respectively, given by
\begin{eqnarray}\label{eq:M}
    \begin{aligned}
        M^2 &=& \Omega^2 +
        \vcenter{\hbox{\includegraphics[width=0.72cm]{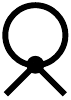}}} +
        \vcenter{\hbox{\includegraphics[width=0.72cm]{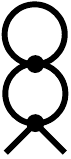}}}
        +
        \vcenter{\hbox{\includegraphics[width=1.3cm]{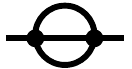}}}
        + \vcenter{\hbox{\includegraphics[width=1.5cm]{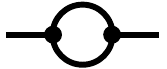}}}\,,
        \qquad
    \end{aligned}
\end{eqnarray}
and
\begin{eqnarray}\label{Z2Gamma}
    \begin{aligned}
        \Gamma^{(1)} &=&-\Omega^2\overline{\chi} +
        \frac{\lambda\overline{\chi}^3}{3} +
        \vcenter{\hbox{\includegraphics[width=0.72cm]{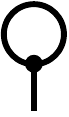}}} +
        \vcenter{\hbox{\includegraphics[width=0.72cm]{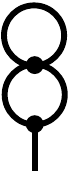}}} +
        \vcenter{\hbox{\includegraphics[width=0.72cm]{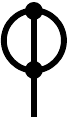}}}\,.
        \qquad
    \end{aligned}
\end{eqnarray}

Evaluating the loop contributions in eqs.~(\ref{eq:M}) and
(\ref{Z2Gamma}) (on mass-shell) within the $\overline{\rm MS}$
dimensional regularization scheme, one obtains,
\begin{eqnarray}
    M^2 &=& \Omega^2 + \frac{\lambda}{8\pi}L_{\Omega} -
    \frac{\lambda^2}{64 \pi^2 \Omega^2} L_{\Omega}
    -\frac{\lambda^2}{384 \Omega^2} -
    \frac{\lambda^2\overline{\chi}^2}{12\sqrt{3}\Omega^2} \,,
\label{M2chiZ2}
  \end{eqnarray}
while $\overline{\chi}$ is fixed by
\begin{equation}
    \Gamma^{(1)} = -\Omega^2\overline{\chi} +
    \frac{\lambda\overline{\chi}^3}{3} -
    \frac{\lambda\overline{\chi}}{8\pi}L_{\Omega}  +
    \frac{\lambda^2\overline{\chi}}{64\pi^2\Omega^2}\overline{\chi}L_{\Omega}
    + \frac{\lambda^2\overline{\chi}}{6(8\pi)^2\Omega^2}C_1\equiv 0
    \,,
\label{chibarZ2}
\end{equation}
where $C_1$ is a constant given by
\begin{equation}
C_1 = \int_0^\infty dr r K_0(r)^3 \simeq 0.586,
\label{C1}
\end{equation}
and $L_{\Omega}$ is defined as
\begin{equation}
    L_{\Omega} = \ln \frac{\mu^2}{\Omega^2}.
\end{equation}

In order to get some numerical results, let us follow other
applications  \cite{Serone:2018gjo,Romatschke:2019rjk,Heymans:2021rqo}
by simply setting $\mu=m$. In this case, when in the subcritical
region, where $\lambda < \lambda_c$ and  ${\overline \chi}=0$, the
pole mass square is given by eq.~(\ref{massaZ2sub1}), while terms
proportional to $L_m$  that appear in there vanish. Hence, tadpoles do
not contribute in the subcritical region. By defining the
dimensionless coupling $g=\lambda/m^2$, eq.~(\ref{massaZ2sub1}) can be
rewritten in  units of $m^2$ as 
\begin{equation}
\frac{M^2}{m^2} = 1  - \frac{g^2}{384  }  \,.
\label{massaZ2sub}
\end{equation}
Since $M^2=0$ at $g=g_c$, one easily finds $g_c = 8\sqrt{6} \simeq
19.59$, which agrees  with the two-loop result found in
refs.~\cite{Serone:2018gjo,Heymans:2021rqo} (once the different
parametrizations represented by a $4!$ factor in the definition of the
$\lambda \phi^4$ interaction have been adjusted). It should be pointed
out that after a resummation including eight-loop perturbative
terms~\cite{Serone:2018gjo,Heymans:2021rqo},  one arrives at the much
higher value $g_c \sim 67$ (when using our normalization). However,
for our present purposes, it is important to notice that the {\it
  qualitative} results here obtained, in particular the order of the
quantum phase transition taking place when the coupling increases, are
not altered by the inclusion of higher order contributions. 

\begin{center}
\begin{figure*}[!htb]
\subfigure[]{\includegraphics[width=7.3cm]{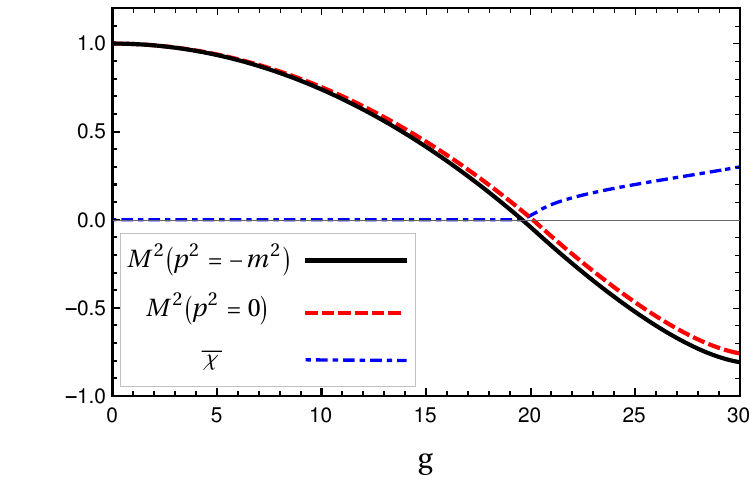}}
\subfigure[]{\includegraphics[width=7.3cm]{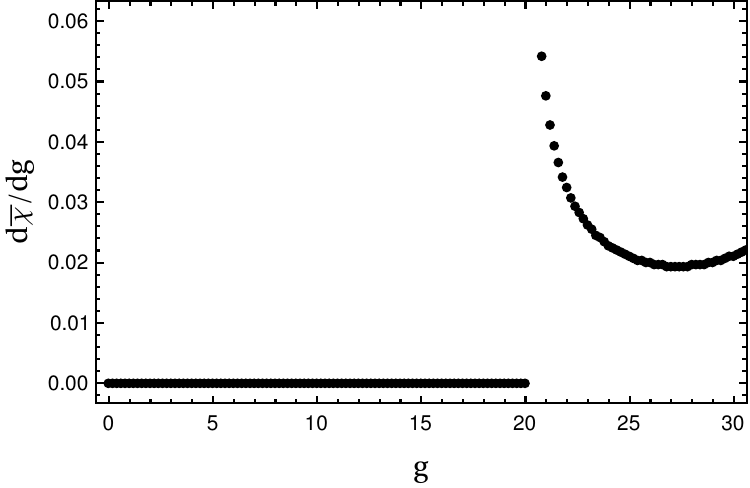} }
 \caption{Panel a: $M^2/m^2$ as a function of $g$  for $M^2(p^2=-m^2)$
   (solid line), $M^2(p^2=0)$ (dashed line) and the VEV
   $\overline{\chi}$  (dash-dotted line). Panel b: the susceptibility
   $d\overline{\chi}/dg$ as a function of $g$.}
 \label{Fig1}
 \end{figure*}
\end{center}

{}Figure~\ref{Fig1}(a) shows the dimensionless mass $M^2/m^2$ and
$\overline{\chi}$ as  functions of  $g$. It illustrates well the
character of the quantum phase transition, as both $M^2/m^2$ and
$\overline{\chi}$ smoothly vanish as the coupling attains its critical
value. {}For completeness, we show in fig.~\ref{Fig1}(a) both the
on-shell square mass $M^2(p^2=-m^2)$ and also the off-shell one,
$M^2(p^2=0)$, where the later is given by 
\begin{equation}
M^2(p^2=0)= m^2 - \frac{\lambda^2}{6(8\pi)^2} C_1 \,.
\end{equation}
At the critical coupling,  the ${\mathbb{Z}_2}$ symmetry is
dynamically broken through a second-order quantum phase transition.
Since $\beta_\lambda =0$ for all coupling values and $M^2(g_c)=0$, the
theory becomes conformal at criticality.  Note that within our
semiclassical loop expansion evaluation, the exact $g_c$ value depends
on whether one uses $M^2(p^2=-m^2)$ or $M^2(p^2=0)$,  but the
difference turns out to be very small anyway ($g_c =8\sqrt{6}\simeq
19.6$ for the former and $g_c \simeq 20.1$  for the later). This small
quantitative difference can be traced back to the fact that we are not
performing any resummations here.  Although   in this first
application to the $O(N_\phi) \times {\mathbb{Z}_2}$ we shall ignore
such  small quantitative discrepancies, we  do expect that by
performing the resummation of high order terms one  will eliminate the
numerical difference observed here. In the same {}fig.~\ref{Fig1}(a)
we also show the VEV $\overline \chi$, which vanishes at the same
point as  $M^2(p^2=0)$.  The fact that both $\overline \chi$ and
$M^2(p^2=0)$ predict the same $g_c$ comes as no surprise, since both
quantities stem from the effective potential (or Landau's free energy)
which generates all 1PI Green's functions with zero external momentum.

The character of the second-order quantum phase transition can also be
verified when computing the VEV susceptibility, $d {\overline \chi}
/dg$, whose behavior is shown in {}fig.~\ref{Fig1}(b) and which
clearly shows the divergence that characterizes a second-order phase
transition happening at $g_c \simeq 20$.  It is worth to point out
that one of the first analysis concerning the quantum phase transition
in the model described by eq.~(\ref{Z2subcritical}) was performed in
ref.~\cite{Chang:1976ek}.  In that reference the Hartree
approximation, which is close to a one-loop analysis including only
the lowest-order diagram in eq.~(\ref{eq:M}), was employed. The result
obtained was however a first-order phase transition, which is in
contradiction with the Simon-Griffiths theorem~\cite{Simon:1973yz}.
Since the model is in the same universality class as the
two-dimensional Ising model, it is then predicted that the transition
should be of second-order~\cite{Simon:1973yz}. Our results, although
simply realized within a two-loop approximation,  do show consistency
with this expectation by correctly predicting the order of the
transition. 

After having shown that  correct {\it qualitative} results regarding
the transition in the case of a single field with a ${\mathbb{Z}_2}$
symmetry can be obtained in the two-loop approximation, let us now
consider the more involved case of a model with  $O(N_\phi) \times
{\mathbb{Z}_2}$ symmetry.  As previously discussed, we are aware that
a (small) numerical difference related to the exact critical coupling
value may arise when the on mass-shell case is not being
resummed. Nevertheless, in the sequel, we shall follow
refs.~\cite{Serone:2018gjo,Romatschke:2019rjk,Heymans:2021rqo} among
many others defining the critical coupling to be the one at which the
on-shell physical mass for a given field vanishes.  {}For our present
purposes, there are two main reasons which justify this
strategy. {}Firstly, our main interest here is, from a qualitative
point of view, to verify not only the transition patterns allowed by
the model, but also  to set the necessary  conditions on the model
parameters such that it will consistently observe the MWHC
theorem. Secondly, we expect that the original perturbative evaluation
of  two-loop terms with on mass-shell contributions to be performed
next  will help to pave the way for future  extensions, including the
resummation of higher order terms.

\section{The $O(N_\phi) \times{\mathbb{Z}_2}$ model}
\label{section3}

Let us now  consider a model composed by two different massive scalar
fields with one of them ($\chi$) having a single component and the
other ($\phi_a$) having $N_\phi$ components. The quartic
self-interactions taking place  within the  different sectors are
parametrized by two different couplings, $\lambda_\chi$ and
$\lambda_\phi$. The two distinct sectors couple via a bi-quadratic
interspecies vertex parametrized by a coupling, $\lambda$. Such
physical system can be conveniently described by an interaction
potential of the form of eq.~(\ref{Vintphichi}) and whose Euclidean
Lagrangian density, invariant under $O(N_\phi)\times{\mathbb{Z}_2}$
global transformations, is given by
\begin{eqnarray}
    \mathcal{L} &=& \frac{1}{2}(\partial_{\mu}\chi)^2 +
    \frac{1}{2}m_{\chi}^2\chi^2 + \frac{\lambda_{\chi}}{4!} \chi^4
    \nonumber \\ &+&  \sum_{a=1}^{N_\phi} \left[
      \frac{1}{2}\partial_{\mu}\phi_a\partial^{\mu}\phi_a +
      \frac{1}{2}m_{\phi}^2\phi_a\phi_a +
      \frac{\lambda_{\phi}}{4!}(\phi_a\phi_a)^2  +
      \frac{\lambda}{4}\phi_a\phi_a\chi^2 \right].
\label{eq:lagranorin}
\end{eqnarray}

Owing to the fact that in this application  we shall be concerned
with the symmetric phase,  the model described by
eq.~(\ref{eq:lagranorin}) will be considered only for the case
$m_i^2 >0$ ($i=\chi,\phi$). It should be recalled that our goal is to
investigate the dynamical quantum phase transition patterns for the
model when it goes from a symmetric to a broken phase, but still
observing the MWHC theorem. As in the previous ${\mathbb{Z}_2}$ case,
the couplings have canonical dimensions
$[\lambda_\phi]=[\lambda_\chi]=[\lambda]=2$, while
$\beta_{\lambda_\phi}= \beta_{\lambda_\chi }= \beta_\lambda \equiv
0$. When the interspecies coupling is negative ($\lambda < 0$), the
boundness condition for the potential requires that the couplings
satisfy the conditions
 \begin{equation}
    \lambda_\phi > 0 \text{ , } \lambda_\chi > 0 \text{ and }
    \lambda^2<\frac{\lambda_\phi\lambda_\chi}{9}\, .
 \end{equation}
  
Since the model described by eq.~(\ref{eq:lagranorin}) has not been
previously considered in $(1+1)$-dimensions, let us briefly discuss
its renormalizability as well as  renormalization group properties.
Just like the related ${\mathbb{Z}_2}$ version the extended
$O(N_\phi)\times {\mathbb{Z}_2}$ version is also super renormalizable
in $(1+1)$-dimensions and the only primitive divergences stem from the
one-loop contributions to the self-energies $\Sigma_i$
($i=\chi,\phi$). Computing the physical masses, $M_i^2 = m_i^2 +
\Sigma_i$, to one-loop order within the $\overline {\rm MS}$
dimensional regularization scheme, one obtains
 \begin{equation} 
 M_\chi^2 = m_\chi^2 + \frac{\lambda_\chi}{8\pi} \left ( \frac
 {1}{\epsilon} + L_{m_\chi} \right ) + \frac{\lambda
   N_\phi}{8\pi}\left ( \frac {1}{\epsilon} + L_{m_\phi} \right ) +
 \delta m^2_{\chi} ,
 \end{equation}
and
\begin{equation} 
 M_\phi^2 = m_\phi^2 + \frac{\lambda_\phi (N_\phi+2)}{24\pi} \left (
 \frac {1}{\epsilon} + L_{m_\phi} \right ) + \frac{\lambda
 }{8\pi}\left ( \frac {1}{\epsilon} + L_{m_\chi} \right ) + \delta
 m^2_{\phi} ,
 \end{equation}
where 
 \begin{equation}
    L_{m_i} \equiv \ln \frac{\mu^2}{m_i^2} \,.
\end{equation}
The contributions $\delta m^2_i$ ($i=\chi,\phi$) represent mass
counterterms whose sole purpose is to  remove the poles.  After
renormalization, each finite physical mass must satisfy the
Callan-Symanzik equation,
\begin{equation}
(\mu \partial_\mu + \beta_{m^2_i} \partial_{m^2_i} + \beta_{m^2_j}
  \partial_{m^2_j}) M^2_i =0 \,,
\end{equation}
from which one obtains
\begin{equation}
\beta_{m_\chi} = - \frac{1}{4\pi} \left ( \lambda_\chi + \lambda
N_\phi \right )\,,
\label{betachi}
\end{equation}
and 
\begin{equation}
\beta_{m_\phi} = - \frac{1}{4\pi} \left [ \lambda_\phi
  \frac{(N_\phi+2)}{3} + \lambda  \right ]\,.
\label{betaphi}
\end{equation}
The above equations generalize the well known ${\mathbb{Z}_2}$ result
given by eq.~(\ref{betam}) to the present case of a $O(N_\phi)\times
{\mathbb{Z}_2}$ invariant model.

\subsection{Physical Masses }

Let us now evaluate the (on mass-shell) physical square masses
considering two regions: (a) the (subcritical) region, where the model
is still $O(N_\phi)\times {\mathbb{Z}_2}$ symmetric and, (b) the
(supercritical) region, where the ${\mathbb{Z}_2}$ symmetry associated
with the $\chi$ sector has been dynamically broken by radiative
corrections. Note that a third (hypercritical) region, where the full
$O(N_\phi) \times {\mathbb{Z}_2}$ symmetry  could be eventually
broken, is excluded by the MWHC theorem. 

\subsubsection{The subcritical region}

{}For the sake of generality, let us consider the case  $m_\phi \neq
m_\chi$. Up to two loops the physical masses receive the following
contributions:
\begin{eqnarray}\label{eq:Veffdia1}
    \begin{aligned}
        M_i^2 &=& m_i^2 +
        \vcenter{\hbox{\includegraphics[width=0.75cm]{tadpole.pdf}}} +
        \vcenter{\hbox{\includegraphics[width=0.75cm]{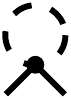}}}
        +
        \vcenter{\hbox{\includegraphics[width=0.75cm]{DoubleScope.pdf}}}
        +
        \vcenter{\hbox{\includegraphics[width=0.75cm]{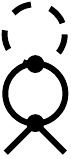}}}
        +
        \vcenter{\hbox{\includegraphics[width=0.75cm]{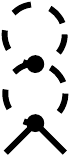}}}
        +
        \vcenter{\hbox{\includegraphics[width=0.75cm]{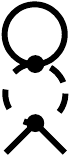}}}
        +
        \vcenter{\hbox{\includegraphics[width=1.4cm]{SettinSun.pdf}}}
        + \vcenter{\hbox{\includegraphics[width=1.4cm]{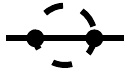}}}
        \,.  \qquad
    \end{aligned}
\end{eqnarray}

Performing the evaluations (see  appendix~\ref{appendixA} for some of
the technical details) within the $\overline {\rm MS}$ dimensional
regularization scheme, one obtains the finite result 
\begin{eqnarray}
    M_i^2 &=& m_i^2 + \frac{\lambda_i (N_i + 2)}{24\pi}L_{m_i} +
    \frac{N_j \lambda}{8\pi}L_{m_j} - \frac{(N_i+2)^2
      \lambda_i^2}{36(4\pi)^2m_i^2}L_{m_i} - \frac{(N_i + 2)N_j
      \lambda_i\lambda}{12(4\pi)^2m_i^2}L_{m_j}  \nonumber \\ &-&
    \frac{(N_j + 2)N_j \lambda_j\lambda}{12(4\pi)^2m_j^2}L_{m_j} -
    \frac{N_iN_j\lambda^2}{4(4\pi)^2m_j^2}L_{m_i}  -
    \frac{(N_i+2)\lambda_i^2}{1152m_i^2} \nonumber \\ &-&
    \frac{N_j\lambda^2}{4(4\pi)^2m_im_j}\left[\pi^2 - 4
      \tanh^{-1}\left(\frac{m_j}{m_i}\right)\ln\left(\frac{m_i}{m_j}\right)
      -
      \frac{m_j}{m_i}\Phi\left(\frac{m_j^2}{m_i^2},2,\frac{1}{2}\right)\right],
\label{subcritMass}
\end{eqnarray}
where $i\neq j = \phi$, $\chi$  and  $N_\chi =1$. In the above
equation $\Phi(z,s,a)$ represents the Lerch transcendent
function~\cite{grads}.

\subsubsection{The supercritical region}

When  the  only phase transition allowed by the MWHC theorem, i.e.,
$O(N_\phi)\times\mathbb{Z}_2 \to O(N_\phi) $,  takes place within the
$\chi$ sector, the Lagrangian density needs to be written in terms of
the shifted field, $\chi^\prime = \chi - {\overline \chi}$. One then
arrives at
\begin{eqnarray}
    \mathcal{L^\prime}
    &=&\frac{1}{2}\partial_{\mu}\phi_a\partial^{\mu}\phi_a +
    \frac{1}{2}\Omega_{\phi}^2\phi_a\phi_a +
    \frac{\lambda_{\phi}}{4!}(\phi_a\phi_a)^2 +
    \frac{1}{2}(\partial_{\mu}\chi^\prime)^2 +
    \frac{1}{2}\Omega_{\chi}^2(\chi^\prime)^2 +
    \frac{\lambda_{\chi}}{4!}(\chi^\prime)^4 +
    \frac{\lambda}{4}\phi_a\phi_a(\chi^\prime)^2 \nonumber \\ &+&
    \frac{\lambda_\chi}{3!}\overline{\chi}(\chi^\prime)^3 +
    \frac{\lambda}{2}\overline{\chi}\phi_a\phi_a\chi^\prime  ,
\end{eqnarray}
where 
\begin{eqnarray}
    \Omega_\phi^2 &=& m_\phi^2 + \frac{\lambda}{2}\overline{\chi}^2
    \,, \\ \Omega_\chi^2 &=& m_\chi^2 +
    \frac{\lambda_\chi}{2}\overline{\chi}^2 .
\end{eqnarray}

To obtain the tadpole equation for $\Gamma^{(1)}$ up to two-loop
order, one needs to evaluate the contributions shown below:
\begin{eqnarray}\label{eq:Veffdia2}
    \begin{aligned}\label{eq:Gamma1}
        \Gamma^{(1)} &=& -\overline{\chi}m_\chi^2 -
        \frac{\lambda_\chi}{6}\overline{\chi}^3 +
        \vcenter{\hbox{\includegraphics[width=0.72cm]{G1-O.pdf}}} +
        \vcenter{\hbox{\includegraphics[width=0.72cm]{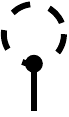}}} +
        \vcenter{\hbox{\includegraphics[width=0.72cm]{G1-OO.pdf}}} +
        \vcenter{\hbox{\includegraphics[width=0.72cm]{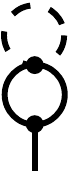}}}
        +
        \vcenter{\hbox{\includegraphics[width=0.72cm]{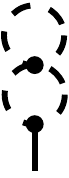}}}
        +
        \vcenter{\hbox{\includegraphics[width=0.72cm]{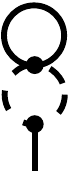}}}
        + \vcenter{\hbox{\includegraphics[width=0.72cm]{G1SSun.pdf}}}
        +
        \vcenter{\hbox{\includegraphics[width=0.72cm]{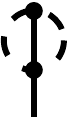}}}
        \,.  \qquad
    \end{aligned}
\end{eqnarray}

Within the supercritical region the masses also receive  new
contributions from the tri-linear vertex which  depends on the VEV,
${\overline \chi}$. This leads to additional terms contributing to
eq.~(\ref{eq:Veffdia1}) which,  for the $\chi$ and $\phi$ fields
respectively, yields

\begin{eqnarray}\label{eq:Mchi}
        M_\chi^2 = \Omega_\chi^2 &+&
        \vcenter{\hbox{\includegraphics[width=0.72cm]{tadpole.pdf}}} +
        \vcenter{\hbox{\includegraphics[width=0.72cm]{tadphichi.pdf}}}
        +
        \vcenter{\hbox{\includegraphics[width=0.72cm]{DoubleScope.pdf}}}
        +
        \vcenter{\hbox{\includegraphics[width=0.72cm]{DDphiphichi.pdf}}}
        +
        \vcenter{\hbox{\includegraphics[width=0.72cm]{DDphichichi.pdf}}}
        +
        \vcenter{\hbox{\includegraphics[width=0.72cm]{DDphichiphi.pdf}}}
        \nonumber \\ &+&
        \vcenter{\hbox{\includegraphics[width=1.3cm]{SettinSun.pdf}}}
        + \vcenter{\hbox{\includegraphics[width=1.3cm]{SSchiphi.pdf}}}
        + \vcenter{\hbox{\includegraphics[width=1.5cm]{YSSun.pdf}}} +
        \vcenter{\hbox{\includegraphics[width=1.5cm]{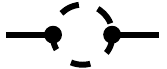}}},
\end{eqnarray}
and
\begin{eqnarray}\label{eq:Mphi}
        M_\phi^2 = \Omega_\phi^2 &+&
        \vcenter{\hbox{\includegraphics[width=0.72cm]{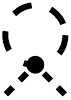}}} +
        \vcenter{\hbox{\includegraphics[width=0.72cm]{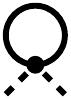}}}
        +
        \vcenter{\hbox{\includegraphics[width=0.72cm]{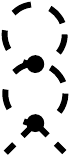}}}
        +
        \vcenter{\hbox{\includegraphics[width=0.72cm]{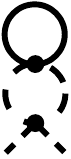}}}
        +
        \vcenter{\hbox{\includegraphics[width=0.72cm]{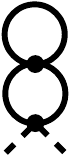}}}
        +
        \vcenter{\hbox{\includegraphics[width=0.72cm]{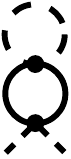}}}
        \nonumber \\ &+&
        \vcenter{\hbox{\includegraphics[width=1.3cm]{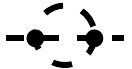}}} +
        \vcenter{\hbox{\includegraphics[width=1.3cm]{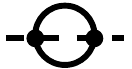}}}
        +
        \vcenter{\hbox{\includegraphics[width=1.5cm]{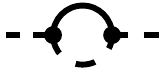}}}.
\end{eqnarray}

Then, the finite results for the masses in eqs.~(\ref{eq:Mchi}) and
(\ref{eq:Mphi}) read
\begin{eqnarray}
    M_\chi^2&=& {\Omega}_{\chi}^2  +
    \frac{\lambda_\chi}{8\pi}L_{\Omega_\chi} +  \frac{N_\phi
      \lambda}{8\pi}L_{\Omega_\phi} - \frac{\lambda_\chi^2}{ (8\pi)^2
      \Omega_\chi^2}L_{\Omega_{\chi}} \nonumber \\  &-&  \frac{N_\phi
      \lambda_\chi \lambda}{(8\pi)^2 \Omega_\chi^2}L_{\Omega_{\phi}} -
    \frac{(N_\phi+2)N_\phi \lambda_\phi \lambda}{3(8\pi)^2
      \Omega_\phi^2 }L_{\Omega_\phi} - \frac{N_\phi
      \lambda^2}{(8\pi)^2 \Omega_\phi^2 }L_{\Omega_\chi} -
    \frac{\lambda_\chi^2}{384 \Omega_\chi^2 } \nonumber
    \\  &-&\frac{N_\phi
      \lambda^2}{(8\pi)^2\sqrt{\Omega_\chi^2\Omega_\phi^2}}\left[\pi^2
      - 2 \,{\rm
        arctanh}\left(\sqrt{\frac{\Omega_\phi^2}{\Omega_\chi^2}}\right)
      \ln\left(\frac{\Omega_\phi^2}{\Omega_\chi^2}\right) -
      \sqrt{\frac{\Omega_\phi^2}{\Omega_\chi^2}}\Phi
      \left(\frac{\Omega_\phi^2}{\Omega_\chi^2},2,\frac{1}{2}\right)\right]
    \nonumber \\ &-&  \frac{\lambda_\chi^2
      \overline{\chi}^2}{12\sqrt{3}\Omega_\chi^2}
    -\frac{N_\phi\lambda^2\sqrt{\pi}\overline{\chi}^2}{8\Omega_\chi^2}
    G^{2{\,}1}_{1{\,}2} \left(
    \left.\frac{4\Omega_\phi^2}{\Omega_\chi^2}\right|\begin{array}{c}
    0,-1/2/,1/2 \\ 0,0,-1/2 
    \end{array} \right),
\label{massaChi}
\end{eqnarray}
where  $G^{2{\,}1}_{1{\,}2} \left( \left. z\right|\begin{array}{c}
0,-1/2/,1/2 \\ 0,0,-1/2 
    \end{array} \right)$ represents the Meijer-G
function~\cite{grads} and
\begin{eqnarray}
   M_\phi^2 &=& \Omega_\phi^2  +
   \frac{(N_\phi+2)\lambda_\phi}{24\pi}L_{\Omega_\phi} +
   \frac{\lambda}{8\pi}L_{\Omega_\chi}  -
   \frac{(N_\phi+2)^2\lambda_\phi^2}{9(8\pi)^2 \Omega_\phi^2
   }L_{\Omega_{\phi}} \nonumber \\  &-&  \frac{(N_\phi+2)\lambda_\phi
     \lambda}{3(8\pi)^2 \Omega_\phi^2 }L_{\Omega_\chi}  -
   \frac{\lambda_\chi \lambda}{(8\pi)^2 \Omega_\chi^2
   }L_{\Omega_{\chi}} -
   \frac{N_\phi\lambda^2}{4(4\pi)^2\Omega_\chi^2}L_{\Omega_{\phi}} -
   \frac{(N_\phi+2)\lambda_\phi^2}{1152 \Omega_\phi^2 } \nonumber
   \\  &-&\frac{\lambda^2}{(8\pi)^2\sqrt{\Omega_\phi^2\Omega_\chi^2}}
   \left[\pi^2 - 2\, {\rm
       arctanh}\left(\sqrt{\frac{\Omega_\chi^2}{\Omega_\phi^2}}\right)
     \ln\left(\frac{\Omega_\chi^2}{\Omega_\phi^2}\right) -
     \sqrt{\frac{\Omega_\chi^2}{\Omega_\phi^2}}\Phi
     \left(\frac{\Omega_\chi^2}{\Omega_\phi^2},2,\frac{1}{2}\right)\right]
   \nonumber \\ &-&
   \frac{\lambda^2\overline{\chi}^2}{2\pi\sqrt{\Omega_\phi^2\Omega_\chi^2}
     \sqrt{4-\frac{\Omega_\chi^2}{\Omega_\phi^2}}}\,{\rm
     arcsec}\left(\sqrt{\frac{4\Omega_\phi^2}{\Omega_\chi^2}} \right)
   ,
\label{massaPhi}
\end{eqnarray}
where
\begin{equation}
  L_{\Omega_i} = \ln \frac{\mu^2}{\Omega_i^2}.
\end{equation}
As before, $\overline \chi$ is determined from $\Gamma^{(1)}=0$, where
\begin{eqnarray}
    \Gamma^{(1)}&=& -\overline{{\chi}}m_\chi^2  -  \frac{\lambda_\chi
      \overline{\chi}^3}{6} - \frac{\lambda_\chi
      \overline{\chi}}{8\pi}L_{\Omega_\chi} -  \frac{N_\phi\lambda
      \overline{\chi}}{8\pi}L_{\Omega_\phi}  \nonumber \\ &+&
    \frac{\lambda_\chi^2\overline{\chi}}{(8\pi)^2\Omega_\chi^2}L_{\Omega_\chi}
    + \frac{N_\phi\lambda^2
      \overline{\chi}}{(8\pi)^2\Omega_\phi^2}L_{\Omega_\chi} +
    \frac{N_\phi(N_\phi+2)\lambda_\phi \lambda
      \overline{\chi}}{192\pi^2\Omega_\phi^2}L_{\Omega_\phi} +
    \frac{N_\phi\lambda_\chi\lambda\overline{\chi}}{(8\pi)^2\Omega_\chi^2}
    L_{\Omega_\phi} \nonumber \\ &+& \frac{\lambda_\chi^2
      \overline{\chi}}{6(8\pi)^2\Omega_\phi^2}C_1 +
    \frac{N_\phi\lambda^2\overline{\chi}}{16\pi^{3/2}\Omega_\chi^2}
    G^{3{\,}2}_{1{\,}0} \left(
    \left.\frac{4\Omega_\phi^2}{\Omega_\chi^2}\right|\begin{array}{c}
    0,0,1/2 \\ 0,0,0,- 
    \end{array} \right).
\label{gamma1}    
\end{eqnarray}

\section{Numerical results}
\label{section4}

{}For numerical evaluations we take $m_\chi= m_\phi \equiv m$, while
defining  the dimensionless couplings $g = \textrm{sgn}(\lambda)
|\lambda|/m^2$, and $g_i = \lambda_i/m^2$, $i=\phi, \, \chi$. Then,
following most applications to the ${\mathbb{Z}_2}$ case, we
set\footnote {Other scales can be chosen in accordance with the masses
  $m_i(\mu)$ running, which is dictated by eqs.~(\ref{betachi}) and
  (\ref{betaphi}).}  the  $\overline {\rm MS}$ scale to  $\mu=m$. With
this particular choice, all tadpole diagrams, which are proportional
to $L_{m_i}$, vanish within the subcritical region. At the same time,
within the supercritical region, the logarithmic terms contained in
the masses and tadpole contributions can be explicitly written as 
\begin{equation}
 L_{\Omega_\chi} = \ln \left( \frac {1}{1+ g_\chi {\overline \chi}^2/2
 } \right) ,
\end{equation}
 and
\begin{equation}
 L_{\Omega_\phi} = \ln \left( \frac {1}{1+g{\overline \chi}^2/2 }
 \right) .
\end{equation}
 
\subsection{Case $O(N_\phi)\times\mathbb{Z}_2$}
 
Applying the definitions  set above for the subcritical masses
described by eq.~(\ref {subcritMass}),  one obtains 
\begin{equation}
  \frac{M_\chi^2}{m^2} = 1- \frac{g_\chi^2}{384} - \frac{g^2}{128}
  N_\phi ,
  \label{mchiSUB}
\end{equation}
and
\begin{equation}
  \frac{M_\phi^2}{m^2} = 1- \frac{g_\phi^2}{384} \frac{(N_\phi+2)}{3}
  - \frac{g^2}{128} .
  \label{mphiSUB}
  \end{equation}
At first glance, one could expect that the symmetry would be
ultimately broken in both channels, however, this is not
guaranteed. The reason is that after the breaking in one direction
(e.g., $\chi$), the masses begin to depend explicitly on the VEV
${\overline \chi}$, as dictated by eqs.~(\ref{massaChi}) and
(\ref{massaPhi}), such that  the supercritical region may be
appropriately described as already mentioned. In this case, $M_\phi^2$
will be affected by $\overline \chi$  as well as by the sign of the
interspecies coupling $g$. In order to investigate whether there is a
possibility that the MWHC theorem will not be violated, it is
instructive to  analyze the phase diagram boundaries on the
$(g_{i,c},\, g)$-plane. {}From eqs.~(\ref {mchiSUB}) and (\ref
{mphiSUB}), we can readily determine the critical couplings
$g_{\chi,c}$ and  $g_{\phi,c}$ as functions of $g$,
\begin{equation}
 g_{\chi,c}(g) = 8\sqrt{6} \left ( 1 - \frac{g^2}{128} N_\phi \right
 )^{1/2} ,
 \label{gcchi}
\end{equation}
and 
\begin{equation}
 g_{\phi,c} (g)= 8\sqrt{6} \left ( \frac{3}{N_\phi +2} \right )^{1/2}
 \left ( 1 - \frac{g^2}{128}  \right )^{1/2}.
 \label{gcphi}
\end{equation}
Imposing that the breaking occurs first in the direction of the $\chi$
field, i.e., $g_{\chi,c}(g) < g_{\phi,c} (g)$,  this implies that $g$
must satisfy the condition
\begin{equation}
g > 8 \sqrt{\frac{2}{N_\phi+3}} .
\label{gchifirst}
\end{equation}
On the other hand, for lower values of $g$, one has instead that
$g_{\phi,c} < g_{\chi,c}$, which means that the $O(N_\phi)$ symmetry
is broken first than the ${\mathbb{Z}_2}$ one violating the MWHC
theorem within this weak coupling regime. As $g$ increases,
$g_{\chi,c}$ approaches $g_{\phi,c}$ until they merge at the  crossing
value $g=g_*$ given by
\begin{equation}
g_* = 8 \sqrt{\frac{2}{N_\phi+3}} .
\end{equation}
After this crossing point we have  $g_{\chi,c} > g_{\phi,c}$ implying
that the ${\mathbb{Z}_2}$ symmetry breaking (in the $\chi$-direction)
occurs first. Once the ${\mathbb{Z}_2}$ symmetry has been broken, the
$\chi$ field acquires a VEV ${\overline \chi}$ and one must use
eqs.~(\ref{massaPhi}) and (\ref{gamma1})  for $M_\phi^2$ and
$\overline \chi$, respectively. Now, the condition $M_\phi^2=0$
determines a new value for $g_{\phi,c}$   which allows us to determine
which region of the parameters space is excluded by the MWHC
theorem. In this situation one may use the two coupled
eqs.~(\ref{massaPhi}) and (\ref{gamma1}) to determine the two unknowns
$g_{\phi,c}$ and $\overline \chi$ for each $g$ value. Since both
equations also depend on $g_\chi$, one can  set $g_\chi = 9
g^2/g_\phi$, which apart from being the threshold for boundness, it
is also exactly satisfied at the crossing point when
$g_{\chi,c}(g_*)=g_{\phi,c}(g_*)$. 

\begin{center}
\begin{figure*}[!htb]
\subfigure[Case $g>0$]{\includegraphics[width=7.3cm]{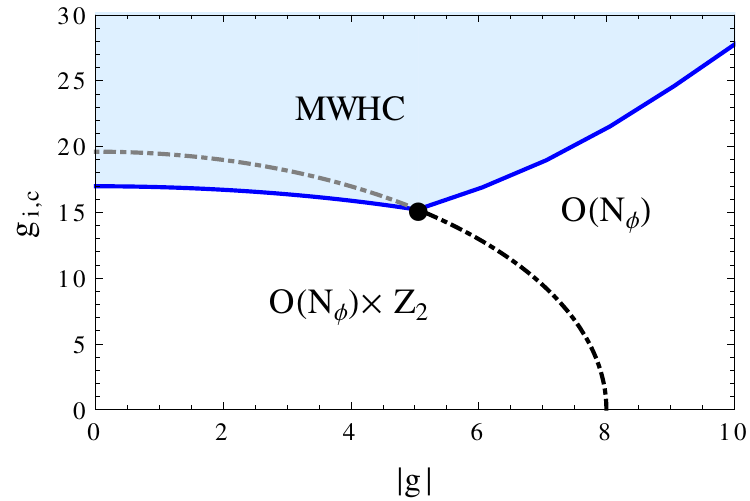}
}  \subfigure[Case
  $g<0$]{\includegraphics[width=7.3cm]{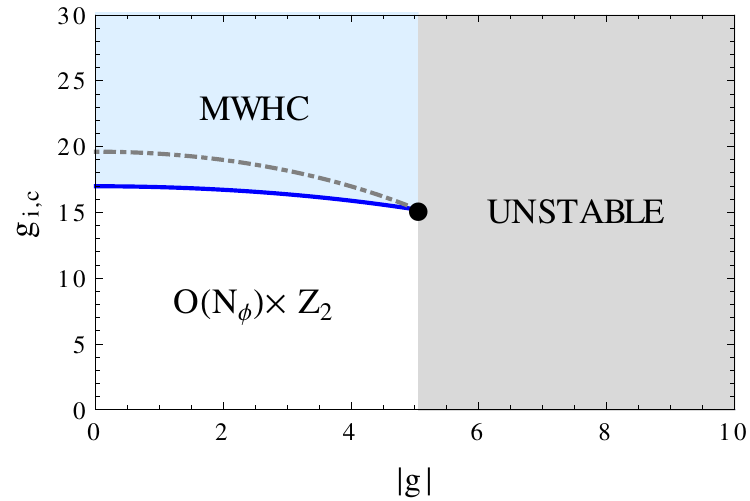} }
 \caption{Phase diagrams on the ($g_{i,c},g$) plane for $g>0$ (panel
   a) and $g <0$ (panel b). The continuous line  represents
   $g_{\phi,c}$ and the dash-dotted line represents $g_{\chi,c}$. In
   both panels the region labelled MWHC indicates where the
   $O(N_\phi)$ symmetry is  broken and thus excluded by the no-go
   theorem. Within the region labelled {\it UNSTABLE} in the right
   panel the boundness condition that applies to the case $g<0$ is not
   observed.  We have taken in this example $N_\phi=2$, which gives
   $g_* = 8 \sqrt{2/5} \simeq 5.05$.}
 \label{Fig2}
 \end{figure*}
\end{center}

In {}fig.~\ref{Fig2} we display the corresponding phase diagrams in
the space of the coupling constants.  The case $g>0$, shown in
fig.~\ref{Fig2}(a), illustrates that it is possible to preserve the
$O(N_\phi)$ symmetry at arbitrarily large $g$ values provided that
$g_\phi < 3g$. The case $g<0$, shown in fig.~\ref{Fig2}(b), describes
a completely different scenario since the theory is only bounded in
the interval $[0,g_*)$ in which the MWHC theorem is violated. To carry
  out the analysis starting exactly at the unstable $g_*$, as we did
  in the $g>0$ case, requires setting $g_\chi = 9 g^2/g_\phi$, which
  violates the boundness condition. As a consequence, the whole region
  $g \ge g_*$ is forbidden. This means that when the couplings are
  independent and the interspecies interaction is repulsive ($g<0$)
  the symmetry breaking pattern $O(N_\phi) \times {\mathbb{Z}_2} \to
  O(N_\phi)$ cannot take place in any regime. {}Finally, we remark
  that $g_*\to 0$ when $N_\phi \to \infty$ so that the MWHC theorem
  covers a region of smaller $g$ values. In the case of $g>0$ this
  means that the breaking $O(N_\phi)\times\mathbb{Z}_2 \to O(N_\phi)$
  can take place at weaker couplings. On the other hand, when $g<0$,
  the first boundary of the unstable region will  start at a smaller
  $g_*$ value  compensating the shrinkage of the MWHC  region and
  forbidding any transition to take place, just like in the $N_\phi=2$
  example shown in {}fig.~\ref{Fig2}.

\begin{center}
\begin{figure*}[!htb]
\subfigure[]{\includegraphics[width=7.3cm]{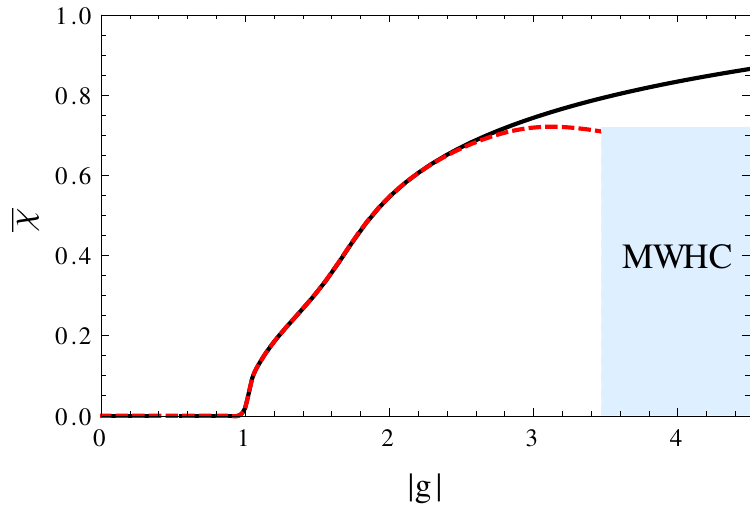} }
\subfigure[]{\includegraphics[width=7.3cm]{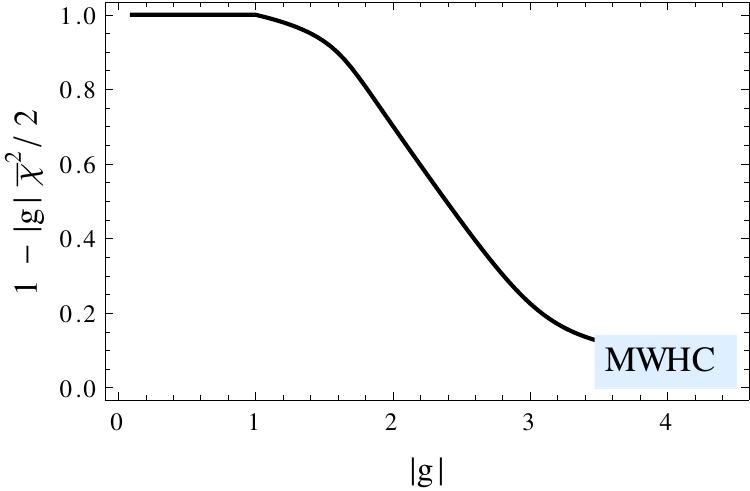} }
 \caption{Panel a: The VEV of the $\chi$ field, $\overline \chi$, as a
   function of $|g|$  for $g>0$ (solid line) and for $g<0$ (dashed
   line). Panel b: $\Omega_\chi^2/m^2$ as a function of $|g|$ in the
   case of $g<0$. The parameters chosen in both plots are $N_\phi = 2$
   while the self-couplings are given by $g_i (g) = k_i |g|$, with
   $k_\chi = 20$ and $k_\phi = 0.5$.}
 \label{Fig3}
 \end{figure*}
\end{center}

It is also useful to explore alternative scenarios where the
self-couplings $g_\phi$ and $g_\chi$ are not independent, but might be
expressed, e.g.,   as a function of the interspecies coupling $g$.
The simplest case is just when the relation between the couplings is
of the form $g_i (g) = k_i |g|$, where the proportionality factor
$k_i$ represents some positive constant. In this case, the breaking of
the ${\mathbb{Z}_2}$ symmetry will certainly occur first if the
self-interactions within the $\chi$ sector are much stronger than the
ones occurring within the $\phi$ counterpart, i.e., $k_\chi \gg
k_\phi$.  To  treat the case where the interspecies coupling is
repulsive ($g<0$), one must further require that $k_\chi k_\phi
>9$. To investigate the different  possible symmetry breaking
patterns, let us suppose for illustration purposes that $k_\chi = 20$
and $k_\phi = 0.5$ as a representative example of parameters
choice. In this case, the requirement that the ${\mathbb{Z}_2}$
symmetry breaks first is satisfied  together with  the boundness
condition. Then, substituting $g_\chi(g)$ and $g_\phi(g)$ respectively
into eqs.~(\ref{gcchi}) and (\ref{gcphi}) one finds that, as expected,
the first breaking occurs in the $\chi$-direction at $g_c = 8
\sqrt{3/203} \simeq 0.97$. To investigate an eventual breaking in the
$\phi$-direction one needs to consider eqs.~(\ref{massaPhi}) and
(\ref{gamma1}) within the supercritical region.  After imposing $\chi$
to be gapless, i.e., $M^2_\chi=0$, one then solves the two coupled
eqs.~(\ref{massaPhi}) and (\ref{gamma1}) to find the two unknowns
$|g_c|$ and $\overline \chi$. Starting with the case $g>0$, one cannot
find a positive solution for $|g_c|$ and $\overline \chi$, whereas
when $g<0$,  one finds $|g_c| \simeq 3.47$ and $\overline \chi \simeq
0.70$. These results corroborate the fact that  the MWHC theorem is
respected (violated) when the interspecies coupling is attractive
(repulsive).  Here, it becomes important to make a digression. Note
that within the supercritical region the VEV ($\overline \chi \ne 0$)
enters  quantities such as $\Omega^2_\phi/m^2= 1 + {\rm sgn}(g) |g|
{\overline \chi}^2 /2$, which  becomes complex if ${\rm sgn}(g)=-$ and
$g {\overline \chi}^2 > 2$. {}The situation is illustrated  by
fig.~\ref{Fig3}(a), which shows ${\overline \chi}$ as a function of
$|g|$ for both cases of ${\rm sgn}(g)=\pm$, and by fig.~\ref{Fig3}(b),
which shows the quantity  $\Omega_\chi^2/m^2= 1 - |g| {\overline
  \chi}^2 /2$ also as a function of $|g|$ when ${\rm sgn}(g)=-$. These
results reassure us that $\Omega_\phi^2 \in \mathbb{R}$ before the
forbidden hypercritical region is reached (up to the point where the
curves reach the MWHC region).

\begin{center}
\begin{figure*}[!htb]
\subfigure[Case $g>0$]{\includegraphics[width=7.3cm]{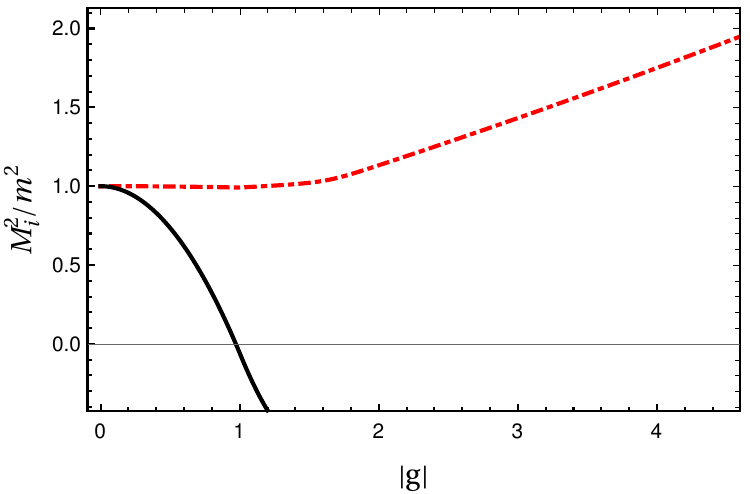} }
\subfigure[Case $g<0$]{\includegraphics[width=7.3cm]{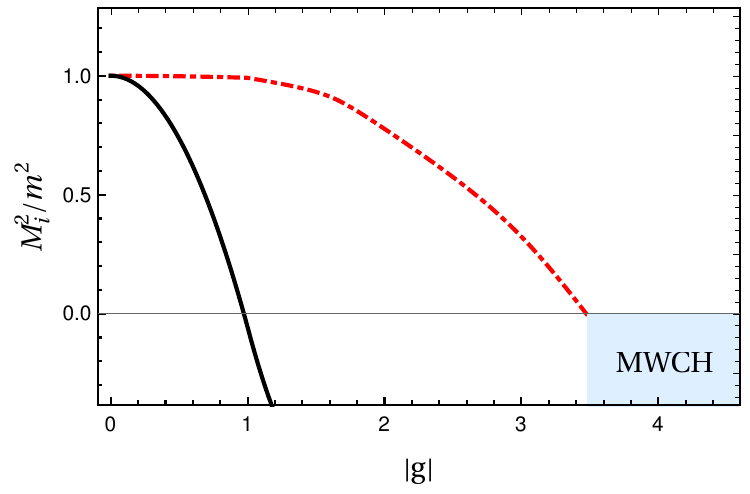} }
 \caption{The pole masses as  functions of $|g|$.  The solid line
   shows $M_\chi^2/m^2$ and the dash-dotted line represents
   $M_\phi^2/m^2$. The same parametrization considered in
   {}fig.~\ref{Fig3} is used here.}
 \label{Fig4}
 \end{figure*}
\end{center}

To get further insight in this case where $g_i = k_i |g|$, let us
investigate the pole masses $M_i^2(g)$ considering ${\rm sgn}(g) =
\pm$ for the case $O(N_\phi) \times {\mathbb{Z}_2}$, with
$N_\phi=2$. The results for this example are shown in
fig.~\ref{Fig4}. They indicate that when $g>0$ the $\phi$ field does
not become gapless as $g$ increases, then respecting the MWHC theorem
at arbitrarily high $g$ values. On the other hand, when $g<0$,
fig.~\ref{Fig4}(b) shows that at $|g_c| \simeq 3.47$ the MWHC region
will be reached and, hence, only the regimes of weak to moderate
coupling values ($|g| < |g_c|$) can be explored.  

In connection to the above results, it is worth to comment on the
validity of the results   in the supercritical regime.    Even though
we do not make use here of a perturbative expansion, but a loop one,
we recall the subtleties of going beyond the gapless phase, which is
intrinsic to  perturbation theory~\cite{Serone:2019szm}. At the
critical point, where $M_\chi^2(g=g_c) =0$, singularities are expected
and $M_\chi^2(g)$ could have different branches.  It is not obvious
that the analytic continuation given by perturbation theory gives the
correct branch. Therefore, following refs.~\cite
{Serone:2018gjo,Serone:2019szm}, we refrain from investigating the
detailed behavior of $M_\chi^2 <0$. {}For our purposes, this course of
action  is totally justifiable since after the first breaking the
order parameter of interest, namely $M_\phi^2$, is affected by
$\overline \chi \ne 0$ rather than  by $M_\chi^2 <0$.

\subsection{Case $\mathbb{Z}_2\times\mathbb{Z}_2$}

{}For completeness, let us now examine the case where $N_\phi=1$ and,
hence, the model is $\mathbb{Z}_2\times \mathbb{Z}_2$ symmetric. Since
now only discrete symmetries are involved, the MWHC theorem does not
apply. {}From eqs.~(\ref{gcchi}) and (\ref{gcphi}), one can see that
in this case the (degenerate) critical couplings are given by
\begin{equation}
 g_{i,c}(g) = 8\sqrt{6} \left ( 1 - \frac{g^2}{128} \right )^{1/2} ,
\label{gdeg}
\end{equation}
which indicates that when the couplings are independent, the
transition $\mathbb{Z}_2\times\mathbb{Z}_2 \to 1$ always happens at
once. 

\begin{center}
\begin{figure*}[!htb]
\subfigure[Case
  $g>0$]{\includegraphics[width=7.3cm]{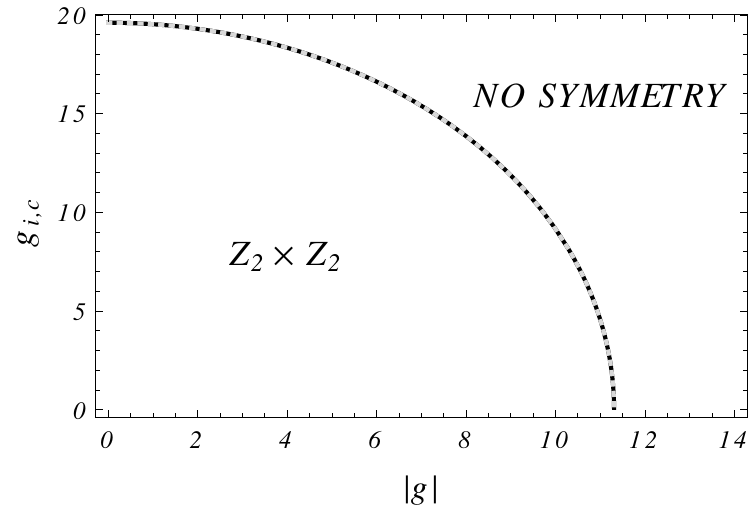} }
\subfigure[Case
  $g<0$]{\includegraphics[width=7.3cm]{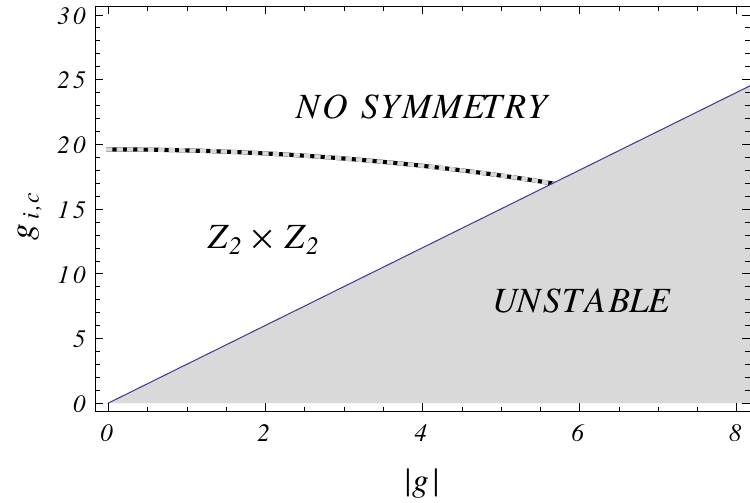} }
 \caption{Phase diagrams in the $g_{i,c},g$ plane for $g>0$ (panel a)
   and $g <0$ (panel b)  for the $\mathbb{Z}_2\times\mathbb{Z}_2$
   symmetric model. The region labelled {\it UNSTABLE} indicates where
   the boundness condition $g_\phi g_\chi > 9 g^2$ is not obeyed. The
   light continuous line representing $g_{\chi,c}$ and the dark dotted
   line  representing  $g_{\phi,c}$ coincide indicating that
   conformability has been attained at the transition boundary.}
 \label{Fig5}
 \end{figure*}
\end{center}

In fig.~\ref{Fig5}(a) we show the resulting phase diagram for
$g>0$. It can be seen that the symmetry will always be completely
broken when the strong coupling regime is attained. The case  $g<0$
requires extra care because although eq.~(\ref{gdeg}) does not depend
on the sign of $g$, the boundness condition still has to be
satisfied. Then, with $g_{\chi,c}=g_{\phi,c}\equiv g_{i,c}$, one must
further impose $g_{i,c} > 3 g$. The phase diagram in this case is
displayed in fig.~\ref {Fig5}(b).   It is important to remark that in
both cases the theory is conformal along the allowed (coincident)
transition lines.

\begin{center}
\begin{figure*}[!htb]
\subfigure[Case $g>0$]{\includegraphics[width=7.3cm]{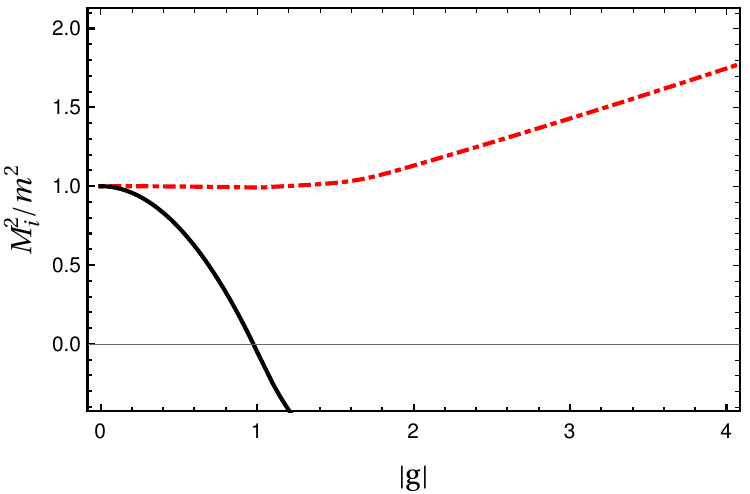} }
\subfigure[Case $g<0$]{\includegraphics[width=7.3cm]{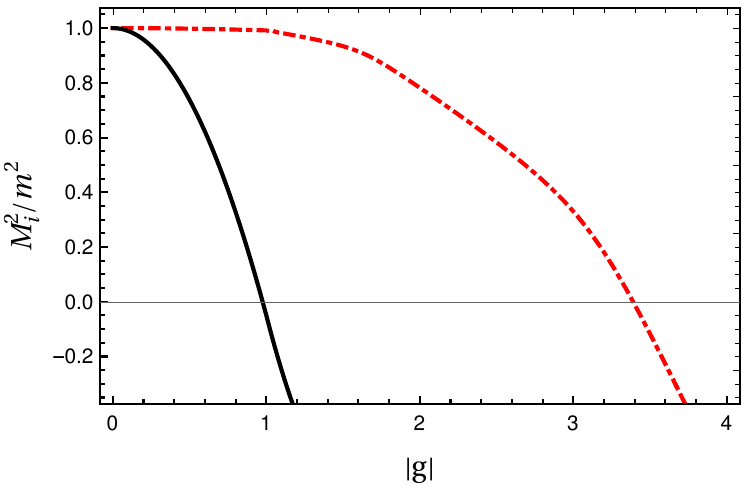} }
 \caption{The pole masses as a function of $|g|$ for the
   $\mathbb{Z}_2\times\mathbb{Z}_2$ symmetric model. The solid line
   shows $M_\chi^2/m^2$ and the dash-dotted line is for
   $M_\phi^2/m^2$. The parameters are chosen susch as $g_{\chi}=20
   |g|$ and $g_{\phi}=0.5 |g|$ in this example.}
 \label{Fig6}
 \end{figure*}
\end{center}

Next, for illustration purposes, let us set $g_{\chi}=20 |g|$ and
$g_{\phi}=0.5|g|$ in order to perform the same type of analysis
carried out in the previous subsection and where the case $O(N_\phi)
\times\mathbb{Z}_2$ was considered. Since now the MWHC theorem does
not apply and the boundness condition is enforced by construction, one
can expected, based on the results obtained for the $O(N_\phi) \times
\mathbb{Z}_2$ case, that two scenarios will emerge: i) when $g>0$ the
transition pattern will be  $\mathbb{Z}_2\times\mathbb{Z}_2 \to
\mathbb{Z}_2 $ and ii) when $g<0$ one expects to also reach  the
hypercritical region, where the remaining  $\mathbb{Z}_2$ symmetry,
associated with the $\phi$ sector, will  be  ultimately broken. {}As
fig.~\ref{Fig6} illustrates,  these  symmetry breaking patterns are
indeed exactly reproduced  according to whether $g>0$
(fig.~\ref{Fig6}(a)) or $g<0$ (fig.~\ref{Fig6}(b)). It is important to
remark that although both sectors, $\chi$ and $\phi$, become gapless
in the $\mathbb{Z}_2\times\mathbb{Z}_2$ case when $g<0$, this does not
automatically imply that the theory will be conformal. As
fig.~\ref{Fig6} reveals, when $g_i = k_i |g|$ and $k_\chi \ne k_\phi$,
the two transitions happen at different coupling values. On the other
hand, as already discussed, when the couplings are independent and the
tree-level masses are degenerate the theory is completely gapless and,
therefore, conformal along the whole transition line displayed in
fig.~\ref{Fig5}.

\subsection{Case $O(N_\phi)\times O(N_\chi)$}

Before closing this section, let us comment on the situation for the
$O(N_\phi)\times O(N_\chi)$ case with $N_i \ge 2$. In this scenario,
the Lagrangian density can be trivially modified with  $\chi$ becoming
a $N_\chi$-component field. In the subcritical region, one would then
obtain the following masses at two-loop order, 
\begin{equation}
  \frac{M_i^2}{m^2} = 1- \frac{g_i^2}{384} \frac{(N_i+2)}{3}-
  \frac{g^2}{128} N_j \,\,\, (i,j=\chi,\phi),
  \label{mchiSUB2}
  \end{equation}
which is the generalization of eqs.~(\ref{gcchi}) and (\ref{gcphi}).
It then happens that the MWHC theorem would be unavoidably violated at
\begin{equation}
 g_{i,c} (g)= 8\sqrt{6} \left ( \frac{3}{N_i +2} \right )^{1/2} \left
 ( 1 - \frac{g^2}{128}N_j  \right )^{1/2}.
 \label{gcphichi}
\end{equation}
Therefore, requiring that none of the continuous symmetries in the
case $O(N_\phi)\times O(N_\chi)$ be broken, which would  violate the
MWHC theorem, sets a bound for the allowed coupling values. Namely,
they must be  sufficiently weak in order to guarantee $M_i^2 > 0$.  On
the other hand, it is worth also recalling that when either $N_\phi
\to \infty$ or $N_\chi \to \infty$, or both are taken in the large-$N$
limit,   Coleman's theorem no longer applies in that specific field
direction~\cite{Coleman:1974jh,Gross:1974jv}. {}Finally, when stable
point-like topological defects can be formed in two dimensional
systems with a spontaneously broken symmetry at zero temperature, we
can have a Berezinskii-Kosterlitz-Thouless (BKT) phase
transition~\cite{Berezinsky:1972rfj,Kosterlitz:1973xp}.  In this case,
a phase transition can occur, yet with no symmetry being truly broken.
This is also a situation where the no-go theorems regarding the
breaking of continuous symmetries at low dimensions can be
evaded. Although this situation could arise in the model  studied
here, which can admit the formation of global topological  nontrivial
solutions, its investigation is far beyond the scope of the present
work.

\section{Conclusions}
\label{conclusions}

We have  analyzed the possible phase transition patterns that may
occur within a bidimensional scalar multi-field model with $O(N_\phi)
\times {\mathbb{Z}_2}$ symmetry. Using the semiclassical loop
approximation to next-to-leading order, we have shown that the
breaking $O(N_\phi) \times {\mathbb{Z}_2} \to O(N_\phi)$, which
respects the Mermin-Wagner-Hohenberg-Coleman no-go theorem may
occur. The transition shows to be of the second kind as in the simpler
${\mathbb{Z}_2}$ version. We have also obtained which conditions the
model parameters need to satisfy in order to observe the no-go
theorem.  {}For instance, when the couplings are independent and  the
interspecies coupling $g$ is attractive ($g>0$),  strong couplings are
allowed, provided that $\mathbb{Z}_2$ breaks first and that $g_\phi <
3 g$. On the other hand, when the couplings are independent and the
interspecies coupling is repulsive ($g<0$), the MWHC theorem together
with the stability condition for the potential, restricts the
couplings to the weak regime. By taking for example the couplings
related through a linear relation, $g_i = k_i |g|$ with $k_\chi \gg
k_\phi$ and $g>0$  the model allows for all values of $g$, while
remaining consistent with the MWHC theorem. On the other hand, if
$g<0$, only   weak and moderate coupling values are allowed.

We have also analyzed the $\mathbb{Z}_2\times \mathbb{Z}_2$ special
case which is not tied up by the MWHC theorem. In this case, the
allowed parameter region is solely restricted when the interspecies
coupling is repulsive, enforcing the stability condition to be
observed. When the couplings are independent, both ($\chi$ and $\phi$)
sectors  become gapless at once, implying  total conformability along
the critical transition line.  {}Finally, we have also discussed the
case with $O(N_\phi)\times O(N_\chi)$ symmetry. In this case, for any
sign of the interspecies coupling, the MWHC theorem restricts the
couplings to  the weak regime only.

In all the cases analyzed, our results indicate that the sign of the
interspecies coupling (attractive/repulsive) can alter the transition
pattern in a significant way.  The results obtained here may be
relevant to  describe the quantum phase transitions taking place in
cold linear systems with competing order parameters, where
descriptions in terms of scalar multi-field systems are
used. Eventually, our model could also serve to investigate
(3+1)-dimensional systems which display a predominance of
(1+1)-dimensional behavior in their physical properties.  An example
of this case can be some variants of the
$XY$-model~\cite{Lima}. A possible extension of the studies performed in this paper could also be applied
in the context of the generalized $O(N)$ nonlinear $\sigma$-model~\cite{abhik}.  Another possibility  would be to check if the large-$N$ and  $\epsilon$-expansion predictions coincide since this comparison furnishes a typical non-trivial consistency check (see, e.g., Ref. \cite {Chai:2020zgq}). {}Finally, we hope that the
super-renormalizable model proposed here will motivate other authors
to test their resummation techniques as well as non-perturbative
methods (such as
\cite{Hauser:1994mb,Rychkov:2014eea,Serone:2018gjo,Romatschke:2019rjk,Kadoh:2018tis,Bronzin:2018tqz,Luty,Chabysheva:2022duu})
in order to improve our seminal results.

\appendix

\section{Evaluation of two loop diagrams with mixed propagators}
\label{appendixA}

Let us present some of the technical details for the evaluation of the
{}Feynman diagrams and which have contribution from both fields, i.e.,
mixed propagators. Working in the $\overline{\textrm{MS}}$ dimensional
regularization scheme, the setting sun like diagram, for instance, is
\begin{eqnarray}\label{eq:ssmisto}
\vcenter{\hbox{\includegraphics[width=2cm]{SSchiphi.pdf}}} &\equiv &
\Sigma_{\rm sun}(p) \nonumber \\ &=& -
\frac{\lambda^2N_j}{4}\left(\frac{\mu^2e^{\gamma_E}}{4\pi}\right)^{\varepsilon}
\int_k\int_q\frac{1}{k^2+m_i^2}\frac{1}{q^2+m_j^2}\frac{1}{(p-k-q)^2 +
  m_j^2}.
\end{eqnarray}
Due to the structure of such diagram, the calculation in the
coordinate space is simpler than in the momentum
space~\cite{Collins:1984xc,Braaten:1995cm}. Writing the propagators in
coordinate space, 
\begin{eqnarray}\label{eq:propx}
    G_i(x) \equiv \mathcal{F}\left\{\frac{1}{l^2 + m_i^2}\right\} =
    \int_l\frac{e^{ilr}}{l^2 + m_i^2} =
    \frac{1}{(2\pi)^{\frac{D}{2}}}\left(\frac{r}{m_i}\right)^{1-\frac{D}{2}}
    K_{1-\frac{D}{2}}(m_ir),
\end{eqnarray}
where $K_i(x)$ is the modified Bessel function of order $i$ allows us
to rewrite eq.~(\ref{eq:ssmisto}) as
\begin{eqnarray}\label{eq:ssmistox}
\Sigma_{\rm sun}(p)= -
\frac{\lambda^2N_\chi}{4}\left(\frac{\mu^2e^{\gamma_E}}{4\pi}\right)^{\varepsilon}
\int\textrm{d}^Dr e^{ipr}G_j^2(x)G_i(x).
\end{eqnarray}
Next, the angular integration in eq.~(\ref{eq:ssmistox}) can be
performed yielding
\begin{equation}\label{eq:ang}
    \int\textrm{d}^Dr e^{ipr} =
    \frac{2\pi^{D/2}}{\Gamma\left(\frac{D}{2}\right)}\int_0^{\infty}\textrm{d}r
    r^{D-1} {}_{0}F_{1}\left(\frac{D}{2},-\frac{p^2r^2}{4}\right),
\end{equation}
with ${}_{i}F_{j}(a,b;c;z)$ representing the hypergeometric
functions. Then, eq.~(\ref{eq:ssmistox}) becomes
\begin{eqnarray}\label{eq:ssx}
\Sigma_{\rm sun}(p)= &=& -
\frac{\lambda^2N_j}{4}\left(\frac{\mu^2e^{\gamma_E}}{4\pi}\right)^{\varepsilon}
\frac{2^{1-\frac{D}{2}}}{(2\pi)^D\Gamma\left(\frac{D}{2}\right)}
\frac{1}{m_j^{2-D}m_i^{1-\frac{D}{2}}}  \nonumber
\\ &\times&\int_0^{\infty}\textrm{d}r
r^{2-\frac{D}{2}}{}_0F_{1}\left(\frac{D}{2},-\frac{p^2r^2}{4}\right)
K^2_{1-\frac{D}{2}}(m_j r)K_{1-\frac{D}{2}}(m_i r)\,.
\end{eqnarray}
{}Finally, by setting $D=2-2\varepsilon$, taking the limit
$\varepsilon \to 0$ and going on mass-shell, $p^2 = -m_i^2$, one
obtains
\begin{eqnarray}
\Sigma_{\rm sun}(p) &=& -
\frac{\lambda^2N_\chi}{4(4\pi)^2m_im_j}\left[\pi^2 -
  4\tanh^{-1}\left(\frac{m_j}{m_i}\right)
  \ln\left(\frac{m_i}{m_j}\right) -
  \frac{m_j}{m_i}\Phi\left(\frac{m_j^2}{m_i^2},2,\frac{1}{2}\right)\right].
\end{eqnarray}
An analogous calculation can be used to obtain the other two-loop
diagrams contributing to the pole masses and to the VEV, $\overline
\chi$.

\acknowledgments

G.O.H. thanks Coordena\c c\~{a}o  de Aperfei\c coamento de Pessoal de
N\'{\i}vel Superior - (CAPES) - Finance  Code  001, for a
Ph.D. scholarship.  M.B.P. is  partially supported by Conselho
Nacional de Desenvolvimento Cient\'{\i}fico e Tecnol\'{o}gico (CNPq),
Grant No  307261/2021-2  and by CAPES - Finance  Code  001.  R.O.R. is
partially supported by research grants from CNPq, Grant
No. 307286/2021-5, and {}Funda\c{c}\~ao Carlos Chagas Filho de Amparo
\`a Pesquisa do Estado do Rio de Janeiro (FAPERJ), Grant
No. E-26/201.150/2021.  This work has also been financed  in  part  by
Instituto  Nacional  de  Ci\^encia  e Tecnologia de F\'{\i}sica
Nuclear e Aplica\c c\~{o}es  (INCT-FNA), Process No.  464898/2014-5.

\providecommand{\href}[2]{#2}\begingroup\raggedright\endgroup

\end{document}